\documentclass[10pt]{article} 

\usepackage[sorting=none,citestyle=numeric-comp]{biblatex}
\usepackage[margin=1in]{geometry}
\usepackage{fancyhdr}
\fancypagestyle{plain}{%
\fancyhf{} 
\fancyfoot[R]{-\,\fontsize{8pt}{8pt}\selectfont\thepage\,-} 

}
\RequirePackage{amsthm,amsmath,amsfonts,amssymb}
\RequirePackage{graphicx, colortbl}
\usepackage{arydshln}

\newcommand{\totem}{\textsc{Tot}\textsc{Em}\xspace}

\usepackage{enumitem}

\usepackage{dsfont}

\usepackage{graphicx}
\usepackage{sidecap}
\usepackage{xspace} 
\usepackage{mathtools}
\usepackage{multirow}
\usepackage{hyperref}

\usepackage{thmtools,thm-restate}

\usepackage{letltxmacro}
\LetLtxMacro{\originaleqref}{\eqref}
\renewcommand{\eqref}{Eq.~\originaleqref}

\newcommand{\multidistro}{\text{mult}\xspace}

\newcommand{\equ}[1]{\begin{gather} #1 \end{gather}}

\newcommand{\quads}[1]{\quad #1 \quad}
\newcommand{\qand}{\quad \text{and} \quad}

\newcommand{\prob}[1]{\mathfrak{#1}}

\makeatletter
\newcommand*\bigcdot{\mathpalette\bigcdot@{.7}}
\newcommand*\bigcdot@[2]{\mathbin{\vcenter{\hbox{\scalebox{#2}{$\m@th#1\bullet$}}}}}
\makeatother

\DeclarePairedDelimiterX\infdivx[2]{(}{)}{
  #1\;\delimsize\|\;#2
}
\DeclarePairedDelimiterX\braket[2]{\langle}{\rangle}{
  #1\;\vert\;#2
}
\DeclarePairedDelimiterX\braketO[3]{\langle}{\rangle}{
  #1\;\vert#2\vert\;#3
}

\newcommand{\infdiv}[2]{D\hspace{0.1mm}\infdivx{#1}{#2}}

\usepackage[colorinlistoftodos,prependcaption,textsize=small]{todonotes}

\definecolor{beaublue}{rgb}{0.74, 0.83, 0.9}
\definecolor{mintcream}{rgb}{0.96, 1.0, 0.98}
\newcommand{\tablecolor}{beaublue}
\definecolor{cadmiumgreen}{rgb}{0.0, 0.42, 0.24}

\begin{document}

\thispagestyle{empty}

\begin{flushright}
\phantom{Version: \today}
\\
\end{flushright}
\vskip .2 cm
\subsection*{}
\begin{center}
{\Large {\bf Eliminating confounder-induced bias \\[1ex] in the statistics of intervention
} }
\\[0pt]

\bigskip
\bigskip {\large
{\bf Orestis Loukas}\footnote{E-mail: orestis.loukas@uni-marburg.de},\,
{\bf Ho Ryun Chung}\footnote{E-mail: ho.chung@uni-marburg.de}\bigskip}\\[0pt]
\vspace{0.23cm}
{\it Institute for Medical Bioinformatics and Biostatistics\\
Philipps-Universität Marburg\\
Hans-Meerwein-Straße 6, 35032 Germany}

\bigskip
\end{center}

\begin{abstract}
\noindent
Experimental and observational studies often lead to spurious association between the outcome and independent variables describing the intervention, because of confounding to third-party factors. 
Even in randomized clinical trials, confounding might be unavoidable due to small sample sizes. 
Practically, this poses a  problem,  because it is either expensive to re-design and conduct a new study or even impossible to alleviate the contribution of some confounders due to e.g.\ ethical concerns. Here, we propose a method to consistently derive hypothetical studies that retain as many of the dependencies in the original study as mathematically possible, while removing any  association of observed confounders to the independent variables. 
Using historic studies, we illustrate how the confounding-free scenario re-estimates the  effect size of the intervention. 
The new effect size estimate represents a concise prediction in the hypothetical scenario which paves a way from the original data towards the design of future studies.
\end{abstract}


\section{Introduction}

A large part of quantitative  literature \cite{blettner1999traditional,valentine2013issues,mathur2019sensitivity,mathur2022methods} in fields ranging from biometrics and clinical studies to sociology and psychometrics is concerned with determining an effect size estimate to quantify the  association of some controlling condition(s) to the outcome of interest. 
Such estimate is usually an $Odds$ ratio ($OR$), relative risk ($RR$) or absolute risk reduction ($ARR$).
Ideally, the computed effect size estimate must offer understanding into the direct causal relationship of the response to an exposure or intervention in the population. This insight is essential for meaningful inference from the data.    

However, various external factors could bias a ``naive'' inference from the study. Consequently, the estimation of  effect size or its direction (e.g.\ risk reduction vs.\ enhancement) might more reflect
some underlying systematic asymmetry in the study population than the targeted relationship between intervention/exposure and outcome.
Conceptualizing the extend of the problem is challenging, as it depends on the chosen theoretical framework and modeling tools used in the statistical analysis. 
The aim of this work is not to offer an overview of the various theoretical and methodological challenges, as e.g.\ discussed in~\cite{cook2002experimental,christenfeld2004risk,brookhart2010confounding}. 
Instead, utilizing a unified framework~\cite{loukas2023demographic} rooted in information theory, we furnish formal tools to eradicate any detected confounding (whether randomly induced or systematic) from the data. 

%
Commonly, clinical trials and observational studies investigate an outcome $y$  represented by a variable $Y$ which could be either categorical like \texttt{success\,\,of\,\,treatment} or metric like \texttt{blood\,\,pressure}. This variable responds to an intervention 
which
comprises the change of some specified conditions summarized by the levels $\boldsymbol x$ of $m$ independent variables  $\boldsymbol X$ like \texttt{treatment\,\,group}, \texttt{diagnostic\,\,tool}, \texttt{geographic location} etc. 
Evidently, $n$
additional variables $\boldsymbol S$ that summarize the demographic, anthropometric or clinical profile $\boldsymbol s$ of studied subjects can influence the outcome $y$. Sometimes such factors are under our control and sometimes not. 
As long as we are not directly interested in the association of $Y$ and $\boldsymbol X$ to $\boldsymbol S$, we can designate  latter variables as (suspected) {confounders} and refer to their influence on the effect size estimate of the intervention as bias. 

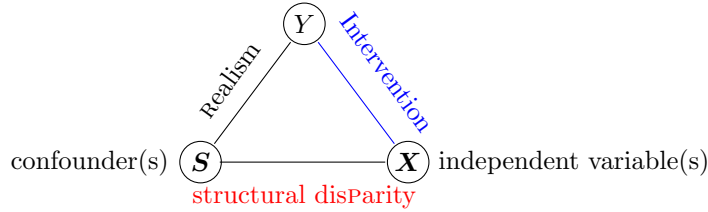
\begin{figure}[!t]
\centering
\begin{tikzpicture}[scale=.92, remember picture ] 
\draw (0,0) circle [radius=0.3] node (Y) {$Y$};
\coordinate (s) at (-1.5,-2);
\draw (s) circle [radius=0.3] node (S) {$\boldsymbol S$};
\draw (s)+(-1.6,0) node {confounder(s)};
\coordinate (x) at (1.5,-2);
\draw (x) circle [radius=0.3] node (X) {$\boldsymbol X$};
\draw (x)+(2.4,0) node {independent variable(s)};
\draw (-1.1,-0.8) node[rotate=54] {\textsc{r}ealism};
\draw (0,-2.5) node[red] {structural dis\textsc{p}arity};
\draw (1.2,-0.7) node[rotate=-54,blue] {Intervention};
\draw[-] (S) -- (Y);
\draw[-] (S) -- (X);
\draw[-,blue] (X) -- (Y);
\end{tikzpicture}
\caption{Undirected three-way graph depicting two-factor associations in the original study.}\label{graph:ThreePoint_disparity}
\end{figure}

The situation resulting from the existence of confounders, which interact with both the outcome $y$ and the examined groups $\boldsymbol x$, is succinctly illustrated in undirected graph~\ref{graph:ThreePoint_disparity}.
In addition, higher-order dependencies might simultaneously associate  all factors leading to non-trivial three-way effects that cannot be predicted just by the drawn edges (hence requiring a hypergraph to depict). 
From contingency tables of the study at hand, the statistical analysis of such graphs on dependencies aims at inferring  an idea about causal associations. Besides asserting a meaningful temporal association (``the cause precedes the outcome''), there is a long way to go to prove causality in the intervention~\cite{greenland1999confounding}, as there might always exist additional confounders that are entirely omitted from the data or when recorded confounders act as a proxy of the actual causal ones. In the following, we concentrate on those attributes $Y$, $\boldsymbol X$ and $\boldsymbol S$ only, that are explicitly  recorded in the study to investigate their associations. 

A first step towards counteracting  the adverse effect of confounders which bias statistical analysis  perplexing causal inference is the --\,often rhetorical\,-- question:
\equ{
\label{hypotheticalQ}
\textit{How would the effect of the intervention on the outcome  look like in a hypothetical scenario}
\nonumber
\\
\textit{where the independent variables would not depend on observed confounders?}
}
There are many reasons to target the particular association of $\boldsymbol X$ to $\boldsymbol S$ in the three-way graph.
At the operational level, structural differences in the distribution of confounders among intervention groups can be manipulated~\cite{vanderweele2019principles} during the design of future experimental and clinical studies. For example, one can decide how many \texttt{smokers\,\,with\,\,asthma} would be included in each \texttt{treatment} group. 

Removing the association between $\boldsymbol X$ and $\boldsymbol S$ balances the generically disparate distribution of confounders over intervention groups $\boldsymbol x$, hence restoring a symmetry~\cite{jaynes2003probability} to enable ``fair'' comparisons among groups. We call this symmetry structural \textsc{p}arity.
Asymptotically given an infinite pool of study subjects, such structural \textsc{p}arity must be exactly realized to manifest the essence of a truly randomized study, at least as far as designated confounders are concerned. Conversely, the expected $\boldsymbol X-\boldsymbol S$ association in a finite study must reflect structural \textsc{p}arity, when the expectation is computed over infinitely many properly randomized trials with finite number of participants each. 

In addition to  conceptual arguments which favor the hypothesis about structural \textsc{p}arity, a cautionary aspect supports targeting this association only:  
Arbitrarily modifying the influence of $\boldsymbol S$ on $Y$ could disrupt the  inherent characteristics of confounders. 
Generally, we  wish to avoid hypothesizing larger modifications that could invalidate biological and physical laws which underlie the associations, unless there is a clear understanding about the association map of $\boldsymbol S$ to $Y$ and $\boldsymbol X$ through external sources. Knowledge transfer is certainly an interesting and fruitful path to explore. For concreteness nevertheless, we henceforth concentrate on the knowledge transmitted by one study individually.

In the present work, we explain how a well-known information-theoretic method can precisely answer the hypothetical question~\ref{hypotheticalQ} by uniquely producing an  alternative expectation about the three-way graph~\ref{graph:ThreePoint_disparity}. In this alternative scenario,  we just remove \textit{one} association in a targeted way, while ensuring that all other dependencies  (including three-factor effects) are left as intact as mathematically possible. 
Without relying on special asymptotic distributions and simplifying assumptions such as equal number of participants in each group or low incidence rates,
this top-down approach offers a unique answer to the  confounding-free scenario 
regarding the dependency of $Y$ on $\boldsymbol X$ that quantifies intervention effects. 
Through various historic paradigms, we exemplify the ability of the suggested methodology to give a definite answer to hypothetical scenario~\ref{hypotheticalQ}  in versatile settings.

\section{Methods}

After collecting  $N$ microdata entries  over a study, we can describe  any subject by specifying the 
three attributes 
$y$, $\boldsymbol x$ and $\boldsymbol s$ from some domains $\mathcal D_Y$, $\mathcal D_{\boldsymbol X}$ and $\mathcal D_{\boldsymbol S}$ respectively.
In that  way, the three-way contingency table can be succinctly provided through the empirical joint distribution $\prob f\equiv \prob f_{Y,\boldsymbol X,\boldsymbol S}$ whose probabilities are given by the relative frequencies $ f(y,\boldsymbol x,\boldsymbol s)$ in the dataset. In the following, we reserve letter $\prob f$ for the empirical statistics of the original study.
Usually, the derived  statistics of a study express our expectation about correlations and associations in the population.
In principle, our expectation for each entry in the three-way contingency table  does not need to adhere to $\prob f$, even if some statistics such as the incidence rate $f_Y(y)$, i.e.\ the prevalence of outcome $y$  are empirically inferred from the given study.

Essentially, the conductor of the analysis is  interested in the interventional  influence  of independent  variables on the outcome, which is represented in the original data by the rate of each outcome $y$ given level $\boldsymbol x$ as a conditional relative frequency
\equ{
\label{eq:InterventionCorrelation}
f_{Y\vert \boldsymbol X}(y\vert \boldsymbol x) = \frac{f_{Y,\boldsymbol X}(y,\boldsymbol x)}{f_{\boldsymbol X}(\boldsymbol x)}
\quads{\text{for}} y\in\mathcal D_Y \qand \boldsymbol x\in\mathcal D_{\boldsymbol X}~.
}
Whenever the additional variables $\boldsymbol S$ become substantially associated to the independent variables $\boldsymbol X$, there might be undesired confounding.
Such non-trivial association between confounders and intervention variables is reflected on the structural difference in the distribution  of confounder profiles $\boldsymbol s$ between the different groups $\boldsymbol x$ quantified by conditional distribution $\prob f_{\boldsymbol S\vert \boldsymbol X}$ with
\equ{
\label{eq:ConfounderDisparity}
f_{\boldsymbol S\vert \boldsymbol X} (\boldsymbol s\vert \boldsymbol x) = \frac{f_{\boldsymbol X,\boldsymbol S}(\boldsymbol x,\boldsymbol s)}{f_{\boldsymbol X}(\boldsymbol x)}
\quads{\text{for}} \boldsymbol  x\in\mathcal D_{\boldsymbol  X} \qand \boldsymbol s\in\mathcal D_{\boldsymbol S}
~,
}
which  will not be generically independent of group $\boldsymbol x$. \eqref{eq:ConfounderDisparity} measures structural heterogeneity~\cite{jager2008confounding,piantadosi2017clinical} in the clinical trial or case-control study; potentially also in the  target population.

More generally, a joint distribution $\prob p$ of $Y$ with $\boldsymbol X$ and $\boldsymbol S$ 
suffices to  unambiguously characterize  all possible associations regarding the attributes like structural differences and confounding effects. 
In particular, any marginal distribution like prevalences $\prob p_Y$, $\prob p_{\boldsymbol X}$ and $\prob p_{\boldsymbol S}$ can be calculated from the joint distribution of the study. 
Furthermore, conditional distributions like $\prob p_{Y\vert X}$ are estimated as ratios of marginals $ p_{Y\vert \boldsymbol X}(y\vert \boldsymbol x) = p_{Y, \boldsymbol X}(y, \boldsymbol x) / p_{\boldsymbol X}(\boldsymbol x)$ from the joint distribution.
Given the conditional distribution %
any odds and risk metric can be inferred.
For our purposes, the domain of attributes over $\mathbb F=\mathbb N,\mathbb Z,\mathbb Q,\{\texttt{true},\texttt{false}\},\dots$ can be empirically determined by
\equ{
\mathcal D_Y = \{y\in\mathbb F \vert f_Y(y)>0\} 
\quads{,} 
\mathcal D_{\boldsymbol X} = \{\boldsymbol x\in\mathbb F^m \vert f_{\boldsymbol X}(\boldsymbol x)>0\} 
\qand
\mathcal D_{\boldsymbol S} = \{\boldsymbol s\in\mathbb F^n \vert f_{\boldsymbol S}(\boldsymbol s)>0\} 
~,
}
so that any distribution $\prob p$ we are going to consider is defined on the simplex over $\mathcal D=\mathcal D_Y \times \mathcal D_{\boldsymbol X}\times\mathcal D_{\boldsymbol S}$. Automatically, this excludes any unobserved profiles $\boldsymbol s$ or non-realized intervention groups $\boldsymbol x$.

Given a distribution $\prob p$ we can sample counts for all elements in the Cartesian product $\mathcal D$. In other words, we sample 
contingency tables, i.e.\ studies, that --\,in expectation\,-- reproduce the statistics of $\prob p$. 
%
In the context of removing confounding effects, any hypothetical scenario is required  to: 
\begin{enumerate}
    \item be self-consistent w.r.t.\ investigated variables $Y,\boldsymbol X$ and $\boldsymbol S$, i.e.\ unequivocally prescribe counts 
    for each cell in the three-way contingency table with all entries summing to $N$ \label{list:wishList1}
    \item respect the observed lower-order marginals \label{list:wishList2}
    \item be confounding-free (at least regarding all designated {observed} confounders)\label{list:wishList3}, i.e.\ $\prob p_{\boldsymbol S\vert\boldsymbol X}=\prob f_{\boldsymbol S}$
    \item otherwise, stay as close as possible to the original statistics of $\prob f$\label{list:wishList4}
\end{enumerate}
The presence of some distribution on the simplex over $\mathcal D$ asserted by point~\ref{list:wishList1} is epistemologically needed in order to be able to discuss the remaining points \ref{list:wishList2},~\ref{list:wishList3} and~\ref{list:wishList4}. 
In that way, the requirement of self-consistency automatically excludes methods like the pooled estimator of the Mantel-Haenszel formula~\cite{10.1093/jnci/22.4.719} that are rooted in heuristics and special (realistically mostly unfulfilled)  limits failing to unambiguously provide a concise estimate for the expectation of the hypothetical contingency table.
As always in statistics, each alternative study on its own does not need to satisfy the last three requirements. Points~\ref{list:wishList2}-\ref{list:wishList4} must become asymptotically fulfilled, when $N\rightarrow\infty$ or by taking the expectation over infinitely many studies at fixed $N$ that are generated from $\prob p$. 
Due to its epistemological importance for the hypothetical study, we refer to $\prob p$ as the fictitious-study distribution, whose $Y-\boldsymbol  X$ marginal distribution logically implies
\equ{
Odds_{\prob p}(\texttt{event}\vert \boldsymbol x) = \frac{\prob p_{Y\vert\boldsymbol  X}(\texttt{event}\vert \boldsymbol x)}{1-\prob p_{Y\vert\boldsymbol  X}(\texttt{event}\vert \boldsymbol  x)}~,
}
leading w.r.t.\ control reference  $\boldsymbol x_0\in\mathcal D_{\boldsymbol X}$  to the  
\equ{
Intervention_{\prob p} = \frac{Odds_{\prob p}(\texttt{event}\vert \,\,\boldsymbol x\,)}{Odds_{\prob p}(\texttt{event}\vert \boldsymbol x_0)}
\quads{\text{for}} \boldsymbol x\in\mathcal D_{\boldsymbol X},
}
as a measure of the effect size in a binary study $\mathcal D_Y = \{\texttt{event}, \texttt{no\,\,event}\}$.
Accordingly, sensible three-factor dependencies can be quantified by 
\equ{
OR = \frac{Odds_{\prob p}(\texttt{event}\vert \boldsymbol x, \boldsymbol s)}{Odds_{\prob p}(\texttt{event}\vert \boldsymbol x_0, \boldsymbol s)}
}
where 
\equ{
Odds_{\prob p}(\texttt{event}\vert \boldsymbol x, \boldsymbol s) = \frac{\prob p_{Y\vert\boldsymbol  X,\boldsymbol S}(\texttt{event}\vert \boldsymbol x, \boldsymbol s)}{1-\prob p_{Y\vert\boldsymbol  X,\boldsymbol S}(\texttt{event}\vert \boldsymbol  x,\boldsymbol s)}
~.
}

The distribution of logistic regression which is often employed~\cite{pourhoseingholi2012control,cepeda2003comparison} in this context fails to fulfill point~\ref{list:wishList3}, since it does not remove structural disparity in the graph~\ref{graph:ThreePoint_disparity}, as we show using the concrete example in Section~\ref{ssc:1948}.
To proceed further we therefore need to employ new tools that we choose from information theory and specifically the framework of \textsc{tot}al \textsc{em}piricism  (\totem~\cite{loukas2023total}).

\subsection{Minimization of information divergence given hypothetical scenario}
\label{ssc:iDiv}

Even if we do not wish to fully adhere to the empirical statistics of $\prob f$, we want any  hypothetical study 
to at least partially  resemble the  statistical properties of the original study, as advocated by point~\ref{list:wishList4}. 
In the spirit of mitigating discriminatory bias~\cite{loukas2023demographic}, the notion of remaining ``close'' to the original contingency table can be covariantly (i.e.\ in a  model-agnostic way) realized as an optimization problem from information theory.
Specifically, we  propose
constructing a joint distribution $\prob p$ as close as possible to the original $\prob f$
that satisfies 
\equ{
\label{eq:PRconditions}
\text{structural \textsc{p}arity:}\quad \prob p_{\boldsymbol X,\boldsymbol S} \overset{!}{=} \prob f_{\boldsymbol X} \cdot \prob f_{\boldsymbol S}
\qand
\text{confounder \textsc{r}ealism}\quad \prob p_{Y,\boldsymbol S}\overset{!}{=} \prob f_{Y,\boldsymbol S}
~.
}

By \textsc{p}arity and \textsc{r}ealism, we exclusively refer to  associations between observed attributes provided  in the original contingency table. Therefore, the present information-theoretic framework is not going to make statements about unobserved confounders~\cite{liu2013introduction,d2019multi}.
As a working assumption of this paper, we further entertain that the observed influence of the confounders on the outcome --\,quantified by marginal $\prob f_{Y,\boldsymbol S}$\,-- fairly depicts reality, at least within the scope of the study.
Instead of fully abiding by \text{confounder \textsc{r}ealism},
one could alternatively replace latter condition with just learning the outcome prevalence  $\prob f_Y$ thus allowing  the optimization task in the next paragraph to infer further details about the influence of suspected confounders on $Y$.

\usetikzlibrary{shapes.misc}
\tikzset{cross/.style={cross out, draw=black, minimum size=2*(#1-\pgflinewidth), inner sep=0pt, outer sep=0pt}, cross/.default={1pt}} 

Signifying by  $\mathfrak{PR}$ the non-empty subset of all  fictitious-study distributions that precisely comply with \eqref{eq:PRconditions}, our objective formally amounts to the determination of a fictitious-study distribution $\prob q\in\mathfrak{PR}$ with the smallest $I$-divergence (also called \textsc{kl}-divergence~\cite{KL_divergence_originalPaper}) from the empirical distribution $\prob f$ describing the original study:
\equ{
\label{eq:Idiv_minimization}
\prob q = \min_{\prob p\in \mathfrak{PR}} \infdiv{\prob p}{\prob f}~.
}
In optimization theory~\cite{nesterov2003introductory}, the former type of conditions \eqref{eq:PRconditions} is called hard constraints and the latter \eqref{eq:Idiv_minimization} soft constraints.
In the flow diagram~\ref{diagram:Parity}, we outline the procedure of finding the fictitious-study distribution as  the $I$-projection of $\prob f$ on $\mathfrak{PR}$:
\begin{figure}[!t]
    \centering
\begin{tikzpicture}[scale=.92, remember picture ] 
\draw (0,0) circle [radius=0.3] node (Y) {$Y$};
\coordinate (s) at (-1.5,-2);
\draw (s) circle [radius=0.3] node (S) {$\boldsymbol S$};
\draw (s)+(-1.6,0) node {confounder(s)};
\coordinate (x) at (1.5,-2);
\draw (x) circle [radius=0.3] node (X) {$\boldsymbol X$};
\draw (x)+(2.4,0) node {independent variable(s)};
\draw (0,-2) node[cross=10pt,red,line width=0.4mm,] {};
\draw (-1.1,-0.8) node[rotate=54] {\textsc{r}ealism};
\draw (0,-2.5) node[red] {structural \textsc{p}arity};
\draw (1.2,-0.7) node[rotate=-54,blue] {Intervention};
\draw[-] (S) -- (Y);
\draw[-] (S) -- (X);
\draw[-,dashed,blue] (X) -- (Y);
\end{tikzpicture}\caption{Undirected three-way graph depicting two-factor associations that are learned from the original study alongside the absence of structural heterogeneity.}\label{diagram:Parity}
\end{figure}
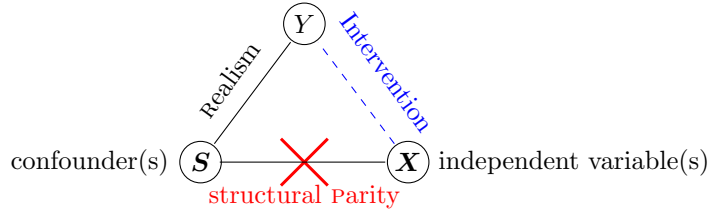
%
Solid lines represent hard constraints imposing an observed association (or its absence), while a dashed line signifies some association that is optimally fixed by the constrained minimization \eqref{eq:Idiv_minimization}. 
Given non-contradictory conditions in \eqref{eq:PRconditions}, it turns out~\cite{csiszar1975divergence} that the joint distribution $\prob q$ with the desired properties, the so-called \textsc{pr}-projection of $\prob f$, is unique. For realistic applications, it can be quickly obtained via numerical methods like \textsc{ipf}~\cite{kruithof1937telefoonverkeersrekening,loukas2022categorical} or covariant Newton-Raphson optimization~\cite{loukas2022entropy,loukas2023total}.

The mathematical guarantee for the convergence of the optimization problem to a distribution $\prob q\in\mathfrak{PR}$ on the simplex over $\mathcal D$
automatically ensures point~\ref{list:wishList1} in our list. 
Using the multinomial distribution $\multidistro(N\prob f_\text{new}\,;\,\prob q)$ we can always sample new studies specifying counts $N f_\text{new}(y,\boldsymbol x, \boldsymbol s)$  for all elements of $\mathcal D$ that would by construction reproduce in expectation the statistics of $\prob q$.
Structural \textsc{p}arity readily ensures then point~\ref{list:wishList3} in the list, while 
conditions \eqref{eq:PRconditions} together imply also point~\ref{list:wishList2}. Finally,  point~\ref{list:wishList4} is also fulfilled in expectation, since repeated sampling  from $\prob q$
concentrates~\cite{Rosenkrantz1989}  around $\prob q$ which is in turn the distribution closest to $\prob f$ under the conditions (in the information-theoretic sense), hence the sampled counts $N\prob f_\text{new}$ are expected to be closest to the original counts $N\prob f$.

Effortlessly, the outlined formalism of $I$-divergence minimization 
can be combined with external sources which update $\prob p_{Y,\boldsymbol S}$ beyond the empirical estimate of $\prob f_{Y,\boldsymbol S}$ on the given data. We will discuss  the option of such knowledge transfer in the future.
In addition, we should not entirely forget higher-order effects (analyzed in biomedical data e.g.\ in~\cite{psf2022005028}) relating the three factors $Y$, $\boldsymbol X$ and $\boldsymbol S$. 
Under the structural homogeneity required by point~\ref{list:wishList3}, such higher-order effects are inferred from the data in the optimal way of \eqref{eq:Idiv_minimization} that ensures the anticipated proximity of $\prob q$  to the original study out of all other distributions in $\mathfrak{PR}$.

Triggered by the imposed absence of structural dis\textsc{p}arity among the groups $\boldsymbol x$, we have a clear expectation  about all entries in the contingency table described by  $\prob q$,
which replaces the observed interventional influence \eqref{eq:InterventionCorrelation} of $\boldsymbol X$ on $Y$ with a hypothetical  
$
\prob q_{Y\vert \boldsymbol X}
$ where by construction \eqref{eq:PRconditions} the conditionals are given by
\equ{
q_{Y\vert \boldsymbol X}(y\vert \boldsymbol x) = \frac{q_{Y,\boldsymbol X}(y,\boldsymbol x)}{f_{\boldsymbol X}(\boldsymbol x)}~.
}
Notice that this is a self-consistent prediction of the hypothetical scenario directly resulting from conditions \eqref{eq:PRconditions} and \eqref{eq:Idiv_minimization}, rather than some fine-tuned or postulated association tormented by (un)intentional bias. As such it offers a glimpse into the expected $Y-\boldsymbol X$ relationship of a truly confounding-free study. 
In addition, the fictitious-study distribution $\prob q$ with the desired features can be used for further statistical analysis. Operating on the combined contingency table (point~\ref{list:wishList1} in the list) enables in a dataset sampled from $\prob q$ a thorough investigation of higher-order associations such as the predictivity~\cite{10.1093/gigascience/giac014}  of profiles $\boldsymbol x$ and $\boldsymbol s$  on the outcome  $y$.

\section{Illustrative examples}

In this Section, we recapitulate four examples from epidemiology and clinical practice in chronological order in order to apply our information-theoretic methodology to remove the influence of designated confounders from the intervention. In most cases, the \textsc{pr}-projection leads to both a qualitative re-assessment of the original conclusions and a quantitative re-estimation of the effect size of intervention on the outcome. 

\subsection{Simpson’s paradox}
\label{ssc:Simpson}

First, we illustrate the information-theoretic methodology based on the historic dataset from~\cite{smith_1934} given in Table~\ref{tb:Simpson}, which is often used as a motivating example for the Yule-Simpson paradox~\cite{yule1903notes,simpson1951interpretation}.
\begin{table}[t]
\renewcommand{\arraystretch}{1.05}
    \centering
    \begin{tabular}{lcl|r||r||r}
         $Y=\texttt{outcome}$ & $X=\texttt{place}$ & $S=\texttt{ethnicity}$ & $N\prob f$ & \textsc{pr}-projection & logit\\
        \hline &&&&&\\[-2.5ex]
        \multirow{4}{*}{ \texttt{survived}} & \multirow{2}{*}{\texttt{New\,\,York}} & \texttt{African\,\,American} & 91\,196 & 0.027413 & 0.018638\\
         && \texttt{White} & 4\,666\,809& 0.944682 & 0.953473\\[0.6ex]
         \rowcolor{\tablecolor}
         & & \texttt{African\,\,American} & 46\,578& 0.000735 & 0.009511\\
         \rowcolor{\tablecolor}
         &\multirow{-2}{*}{\texttt{Richmond}}  & \texttt{White} & 80\,764&0.025297 & 0.016506\\
         \hdashline &&&&&\\[-2ex]
         \multirow{4}{*}{ \texttt{died}} & \multirow{2}{*}{\texttt{New\,\,York}} & \texttt{African\,\,American} & 513 &0.000134 & 0.000100\\
         && \texttt{White} & 8\,365& 0.001695 & 0.001714\\[0.6ex]
         \rowcolor{\tablecolor}
         &  & \texttt{African\,\,American} & 155& 0.000002 & 0.000037\\
         \rowcolor{\tablecolor}
         &\multirow{-2}{*}{\texttt{Richmond}} & \texttt{White} & 131 & 0.000041 & 0.000022\\
    \end{tabular}
    \caption{Historic data for the tuberculosis pandemic in the \textsc{us}. Besides the observed counts $N\prob f$, we provide the probabilities of the \textsc{pr}-projection of $\prob f$ and of logistic regression on the data for all cells in the three-way contingency table.}
    \label{tb:Simpson}
\end{table}
In this study from 1910, we investigate the mortality rate of tuberculosis observed due to disparate living conditions in 
different areas of the country as experienced by $N=4\,894\,511$ individuals. The intervention is thus related to the $\texttt{place}$ of residence with the associated  influence on mortality quantified by 
\equ{
f_{Y\vert X} (\texttt{died}\vert \texttt{place}) \quads{\text{with}} \texttt{place}\in\{\texttt{New\,\,York}, \texttt{Richmond}\}
\nonumber
}
empirically resulting into 
\equ{
\label{eq:Simpson:target_OR}
Intervention_{\prob f} = \frac{Odds_{\prob f}(\texttt{died}\vert\texttt{Richmond})}{Odds_{\prob f}(\texttt{died}\vert\texttt{New\,\,York})} \approx 1.2
}
and hence significantly ($p\text{-value}\approx0.0025$ by Fisher's exact test) pointing towards an increased risk in \texttt{Richmond}. 
At first, this might seem unexpected (in other words ``paradoxical''), due to the lower population density in \texttt{Richmond} as opposed to \texttt{New\,\,York}, which --\,if anything\,-- should reduce infection and hence mortality rates. 

Designating \texttt{ethnicity} as a potential confounder sheds  some light into the ``paradoxon''. 
Indeed, a three-point analysis stratified by the two ethnic groups gives 
\equ{
\label{eq:Simpson:Stratification}
\frac{Odds(\texttt{died}\vert\texttt{Richmond},\texttt{African\,\,American})}{Odds(\texttt{died}\vert\texttt{New\,\,York},\texttt{African\,\,American})} \approx 0.59
\quads{\text{vs.}}
\frac{Odds(\texttt{died}\vert\texttt{Richmond},\texttt{White})}{Odds(\texttt{died}\vert\texttt{New\,\,York},\texttt{White})} \approx 0.90
}
hinting towards the more intuitive effect of lower population density to prohibit the spread of tuberculosis. 
Therefore, \texttt{ethnicity} acts as a significant ($p\text{-value}\approx0$) confounder due to the \textsc{r}ealism of
\equ{
\label{eq:Simpson:socioEconomicRealism}
\frac{ Odds(\texttt{died}\vert \texttt{African\,\,American})}{Odds(\texttt{died}\vert \texttt{White})\phantom{...................}}
\approx
2.71~,
}
presumably reflecting the socioeconomic situation at that time. 
This contributes 
together with significantly ($p\text{-value}\approx0$) unbalanced representation of considered minorities among the regions of interest,
\equ{
\label{eq:Simpson:StructuralDifferene}
p_{S|X}(\texttt{African\,\,American}\vert \texttt{New\,\,York}) \approx 2\% 
\quads{\text{vs.}}
p_{S|X}(\texttt{African\,\,American}\vert \texttt{Richmond}) \approx 37\% 
~,
}
to make \texttt{Richmond} appear more dangerous in \eqref{eq:Simpson:target_OR}.
The stratified presentation of 
\eqref{eq:Simpson:Stratification} has  not yet removed the confounding to \texttt{ethnicity} of the intervention in the \textit{combined} population of Table~\ref{tb:Simpson}.

In fact, the two $OR$'s given \texttt{ethnicity} are considerably different rendering further analysis on the original dataset and  interpretation challenging.
As already noted in the original paper~\cite{10.1093/jnci/22.4.719}, a pooled risk\footnote{Even in this application with small mortality rates below $1\%$, the Mantel-Haenszel method computes a $OR_{\text{pool}}\approx0.73$ which overestimates a confounder-free intervention, as we are going to see by the information-theoretic methods which give a higher $OR$.} out of the two strata does not clearly prescribe a combined, three-way contingency table, as it is motivated by arguments (except perhaps in limiting scenarios~\cite{TARONE1981463,NomaNagashima_2016_19_35}) on weighing schemes according to the relative importance of the stratified risks.
Therefore, the  Mantel–Haenszel estimator does not mediate a clear idea of how a confounding-free study (or any hypothetical study that exhibits such pooled $OR$) should look like, since the reasoning behind it does not necessarily fulfill point~\ref{list:wishList1}.

The fictitious-study distribution $\prob q$  incorporating conditions~\eqref{eq:PRconditions} eliminates the structural difference by the global prevalence of the minority (as extracted from the study data)  thus replacing \eqref{eq:Simpson:StructuralDifferene} with
\begin{align}
\label{eq:StructuralParity}
q_{S|X}(\texttt{African\,\,American}\vert \texttt{New\,\,York}) =
q_{S|X}(\texttt{African\,\,American}\vert \texttt{Richmond})& 
\\
=&\,\, f_S(\texttt{African\,\,American}) \approx 2.8\%
\nonumber
\end{align}
while fully adhering to the socioeconomic realism of \eqref{eq:Simpson:socioEconomicRealism}.
As a result of $I$-divergence minimization, the \textsc{pr}-projection re-estimates  
\equ{
\label{eq:Simpson:Intervention_PRprojection}
Intervention_{\prob q} = \frac{Odds_{\prob q}(\texttt{died}\vert\texttt{Richmond})}{Odds_{\prob q}(\texttt{died}\vert\texttt{New\,\,York})} \approx 0.882~,
}
which implies that the  population in \texttt{Richmond} had a reduced risk compared to \texttt{New\,\,York}, as anticipated by the stratified analysis of \eqref{eq:Simpson:Stratification}. 
We use subscript ${\prob q}$ to signify that those $Odds$ are computed using probability estimates of the \textsc{pr}-projection in Table~\ref{tb:Simpson}.
Incidentally, removing the hard condition of confounder Realism in \eqref{eq:PRconditions}, i.e.\
only learning 
the hard constraints of confounder \textsc{p}arity and global death-rate prevalence, leads approximately to the same intervention $OR\approx0.879$. 
In other settings however, the two approaches to learning confounder \textsc{r}ealism via soft as opposed to hard conditions do not need to even approximately agree.

Going beyond the stratified analysis, \eqref{eq:Simpson:Intervention_PRprojection} directly reports on one, --\, the desired\,-- $OR$ to self-consistently quantify the effect of intervention regarding \texttt{place} of residence in the hypothetical scenario of complete \textsc{p}arity in the representation of \texttt{ethnicities}  between the two cities. 
At this point, one might be tempted to sample new (synthetic in the jargon of machine learning) data from the \textsc{pr}-projection to compute the $p$-value associated to the predicted interventional effect size in \eqref{eq:Simpson:Intervention_PRprojection}. Due to the statistical nature of our results, which only remove structural heterogeneity \textit{in expectation}, and  given  that the scenario is at the outset hypothetical, we refrain from quoting $p$-values related to $\prob q$. Instead, we suggest to interpret the information-theoretic analysis --\,beyond a mere explanation of confounding paradoxes\,-- as a prediction about alternative studies to direct future research along the lines of \eqref{eq:PRconditions}. 

After establishing \textsc{p}arity in the representation of  \texttt{ethnicities} between \texttt{places}, our expectation $\prob q$ about three-way Table~\ref{tb:Simpson}
has now changed. The information-theoretic argument of \eqref{eq:Idiv_minimization} guarantees that this change reflects the minimal (in the sense of $I$-divergence minimization~\cite{shore1980axiomatic,csiszar1991least}) modification of probabilities in the contingency table that is needed to ensure 
 the same distribution of designated confounder (\texttt{ethnicity}) in both groups (\texttt{place}).
 In particular, out of all three-way metrics  the stratified $OR$ by \texttt{ethnicity} retain their empirical values given in \eqref{eq:Simpson:Stratification}:
 \equ{
\frac{Odds_{\prob q}(\texttt{died}\vert\texttt{Richmond},\texttt{African\,\,American})}{Odds_{\prob q}(\texttt{died}\vert\texttt{New\,\,York},\texttt{African\,\,American})} \approx 0.59
\quads{\text{vs.}}
\frac{Odds_{\prob q}(\texttt{died}\vert\texttt{Richmond},\texttt{White})}{Odds_{\prob q}(\texttt{died}\vert\texttt{New\,\,York},\texttt{White})} \approx 0.90
}
In summary, our hypothetical scenario with structural \textsc{p}arity \eqref{eq:StructuralParity} self-consistently reproduces the empirical stratified analysis complementing it with the interventional $OR$ of \eqref{eq:Simpson:Intervention_PRprojection}. 

Another approach to addressing confounding might involve modeling $\prob f$. Specifically, we apply conventional logistic regression (logit) which minimizes~\cite{loukas2023total} the $I$-divergence from the uniform distribution   (conversely maximizes~\cite{jaynes2003probability} the entropy) under the hard constraints 
\equ{
\label{eq:Simpson:Logit}
\text{logit:}\quad
\prob p_{\boldsymbol X,\boldsymbol S} \overset{!}{=}\prob f_{\boldsymbol X, \boldsymbol S}
\quads{,}
\prob p_{ Y, \boldsymbol X} \overset{!}{=}\prob f_{ Y, \boldsymbol X}
\qand
\prob p_{ Y,  \boldsymbol S} \overset{!}{=}\prob f_{ Y, \boldsymbol S}~.
}
Actually, this set of pairwise associations would be generically counterfactual, since they neglect any 
higher-order effects.
Computing 
the three-point ratio
\equ{
\label{eq:Simpson:Logit_ThreePoint_OR}
\frac{Odds_{\prob l}(\texttt{died}\vert\texttt{Richmond},\texttt{White})}{Odds_{\prob l}(\texttt{died}\vert\texttt{New\,\,York},\texttt{White})} =
\frac{Odds_{\prob l}(\texttt{died}\vert\texttt{Richmond},\texttt{African\,\,American})}{Odds_{\prob l}(\texttt{died}\vert\texttt{New\,\,York},\texttt{African\,\,American})} \approx 0.726
~,
}
where ${\prob l}$ denotes the joint distribution\footnote{In the parametric approach to logistic regression, the parametrized conditional distribution $\prob l_{Y|\boldsymbol X,\boldsymbol S}$ trained on $\prob f$ can be always applied on the training profiles distributed according to $\prob f_{\boldsymbol X,\boldsymbol S}$, so that the joint distribution $l(y,\boldsymbol x,\boldsymbol s)= l_{Y|\boldsymbol X,\boldsymbol S}(y\vert \boldsymbol x,\boldsymbol s)\cdot f_{\boldsymbol X,\boldsymbol S}(\boldsymbol x,\boldsymbol s)$ describes the expectation of trained logit, once asked to predict on the training data itself.} of logistic regression under \eqref{eq:Simpson:Logit} given in the last column of Table~\ref{tb:Simpson}, shows that --\,by construction\,-- the discrepancy of the stratified analysis \eqref{eq:Simpson:Stratification} is now gone. However, the logit scenario  neither deals with the actual problem, nor shows how to self-consistently resolve it, as the two-point ratio quantifying the effect size of the intervention 
\equ{
OR_{\prob l} = \frac{Odds_{\prob l}(\texttt{died}\vert\texttt{Richmond})}{Odds_{\prob l}(\texttt{died}\vert\texttt{New\,\,York})} = OR_{\prob f}\approx 1.20 
}
still retains --\,also by construction\,-- its empirical value. In fact, the three-point $OR$ in \eqref{eq:Simpson:Logit_ThreePoint_OR} appears to be inconsistent with \eqref{eq:Simpson:Intervention_PRprojection}, which estimates the effect size for the intervention of interest in the hypothetical scenario with no structural heterogeneity that still remains closest to the original study (in the information-theoretic sense), as  we have argued above.

\subsection{First randomized clinical trial}
\label{ssc:1948}

\begin{table}[t]
\renewcommand{\arraystretch}{1.05}
    \centering
    \begin{tabular}{rll|ll}
    \multicolumn{2}{c}{$\boldsymbol S$\phantom{...}} & $Y$ & \multicolumn{2}{c}{$X$\phantom{.....}}
    \\ 
         \texttt{gender} & \texttt{baseline} & $\texttt{improved}$ & \texttt{control} & \texttt{streptomycin} \\ \hline &&&&\\[-2.5ex]
                        &  & \texttt{no} & \phantom{0}0 (0) & \phantom{0}0 (0)\\
                        &  \multirow{-2}{*}{\texttt{good}}  & \texttt{yes} & \phantom{0}4 (0.036335) & \phantom{0}4 (0.038431)\\
        \rowcolor{\tablecolor}                
         & & \texttt{no} & \phantom{0}5 (0.045886) & \phantom{0}2 (0.019534) \\
        \rowcolor{\tablecolor}
        \multirow{-2}{*}{\texttt{female}} &  \multirow{-2}{*}{\texttt{fair}} & \texttt{yes} & \phantom{0}5 (0.044951) & \phantom{0}8 (0.076544)\\
                        & & \texttt{no} & 14 (0.140798) & 10 (0.083501)\\ 
                        &  \multirow{-2}{*}{\texttt{poor}}  & \texttt{yes} & \phantom{0}0 (0) & \phantom{0}7 (0.065421)\\[.5ex]
        \hdashline&&&&\\[-2ex]
        \rowcolor{\tablecolor} 
                        & & \texttt{no} & \phantom{0}0 (0) & \phantom{0}0 (0)\\
        \rowcolor{\tablecolor} 
                        &  \multirow{-2}{*}{\texttt{good}}  & \texttt{yes} & \phantom{0}4 (0.036335) & \phantom{0}4 (0.038431)\\ 
                        &  & \texttt{no} & \phantom{0}6 (0.050943) & \phantom{0}1 (0.014478) \\
        \multirow{-2}{*}{\texttt{male}}  & \multirow{-2}{*}{\texttt{fair}} & \texttt{yes} & \phantom{0}4 (0.026269) & \phantom{0}6 (0.067189)\\ 
        \rowcolor{\tablecolor} 
                        &  & \texttt{no} & 10 (0.104463) & \phantom{0}4 (0.026378)\\ 
        \rowcolor{\tablecolor} 
                        & \multirow{-2}{*}{\texttt{poor}} & \texttt{yes} & \phantom{0}0 (0) & \phantom{0}9 (0.084112)\\
                        
    \end{tabular}
    \caption{Controlled investigation performed in 1948 with $N=107$ patients into the effects of the antibiotics streptomycin on one
type of pulmonary tuberculosis. For each patient profile and possible outcome, we provide the observed counts and  the probability estimates of \textsc{pr}-projection (parenthesis) in the two treatment groups.}
    \label{tb:Clinical1948}
\end{table}

Next, we turn to a paradigm~\cite{Doll1217} for the ``golden-rule'' of randomized clinical trials, a systematic clinical trial 
on $X=\texttt{treatment}$ 
which incorporated the contemporary ideas of randomization and  concealment of allocation advocated by~\cite{Hill1043}. Colloquially, this trial is regarded as the first randomized clinical trial in modern era. 
From a scanned version of the original study~\cite{original1948},
we reconstructed three-way contingency Table~\ref{tb:Clinical1948} using the \textsc{r}-package~\cite{medicaldata}.
In the trial, 
$Y=\texttt{improvement}$ of the clinical condition of patients towards recovery was assessed at multiple endpoints such as changes in the radiological picture, weight and temperature over the trial period. The \texttt{control} therapy consisted of the only known~\cite{doi:10.1177/014107680609901017} reference method at the time --\,bed rest, while the experimental therapy comprised the administration of the antibiotics \texttt{streptomycin} alongside bed rest. 

In our investigation, a binary attribute (\texttt{gender}) and a categorical attribute (\texttt{baseline} condition)  are both designated as potential  confounders $\boldsymbol S$. The  information-theoretic formalism of $I$-divergence minimization allows to simultaneously eliminate  any structural difference described by linear as well as non-linear effects. In that way, all \texttt{treatment} groups $\mathcal D_X=\{\texttt{control},\texttt{streptomycin}\}$ would exhibit the same conditional distribution $\prob f_{\boldsymbol S\vert X}$ of the two confounders over the six possible manifestations (first two columns of Table~\ref{tb:Clinical1948})
\equ{
\label{eq:1948:s-Profiles}
\mathcal D_{\boldsymbol S} = \{ (\texttt{female,\,\,\texttt{good}}), (\texttt{female,\,\,\texttt{fair}}),(\texttt{female,\,\,\texttt{poor}}),(\texttt{male,\,\,\texttt{good}}),(\texttt{male,\,\,\texttt{fair}}), (\texttt{male,\,\,\texttt{poor}}) \}
~.
}
Despite the presence of unobserved profiles (so-called sampling zeros~\cite{bishop2007discrete}) in the contingency table, the marginal distribution $\prob f_{X,\boldsymbol S}$ remains strict positive. If  some patient profile $\boldsymbol s$ were not observed in both \texttt{treatment} groups, either regularization or modeling would be warranted\footnote{A (pre-)analysis of finite (and small) sample-size  effects, which in any case undermines statistical power, goes beyond the scopes of the present paper.}. 

Given the $N=107$ patients, we can empirically quantify the  effect of 
\equ{
\label{eq:1948:Intervention}
Intervention_{\prob f} = \frac{Odds_{\prob f}(\texttt{improved} \vert \texttt{streptomycin})}{Odds_{\prob f}(\texttt{improved} \vert \texttt{control})\phantom{.........}} \approx 4.60
}
corresponding to an absolute risk reduction
of more than $36\%$ 
which results in $p\text{-value}\approx 0.0002$. Because of the relatively small study and broader concerns~\cite{rosenberger2015randomization,https://doi.org/10.1002/sim.4780060325} regarding ethics and study design, one might wonder whether this significantly high drug effectiveness  is not an artifact of confounding. 
Structural heterogeneity among the \texttt{treatment} groups can be quantified  via the ratio of conditionals
\equ{
\label{eq:1948:StructuralHeterogeneity}
\frac{ f_{\boldsymbol S\vert X}(\boldsymbol s\vert\texttt{streptomycin})}{ f_{\boldsymbol S\vert X}(\boldsymbol s\vert\texttt{control})\phantom{.........}} \approx\,
\begin{array}{l|ccc}
    \renewcommand{\arraystretch}{1.1}
        & \texttt{good} & \texttt{fair} & \texttt{poor} \\\cline{2-4}
        \texttt{Female} & 0.95 & 0.95 & 1.15 \\
        \texttt{Male}   &  0.95 & 0.66 &  1.23
    \end{array}
    ~,
}
which demonstrates that  the \texttt{streptomycin} group has been disfavoured with more cases of \texttt{poor} baseline condition  for both \texttt{female} and \texttt{male}.
This could be problematic regarding the effect size estimate \eqref{eq:1948:Intervention}, since different preconditions 
are anticipated to largely support or undermine recovery from disease. 
Indeed, the \texttt{baseline} condition for example has a significantly strong impact on the outcome
\equ{
\frac{Odds_{\prob f}(\texttt{non-improved}\vert \texttt{fair})}{Odds_{\prob f}(\texttt{non-improved}\vert \texttt{poor})}\approx 0.26
\qand
\frac{Odds_{\prob f}(\texttt{non-improved}\vert \texttt{good})}{Odds_{\prob f}(\texttt{non-improved}\vert \texttt{poor})}=0
}
with $p\text{-value}\approx0.003$ and $p\text{-value}\approx0.015$, respectively.

\begin{figure}[!t]
    \centering
    \includegraphics[clip, trim=31mm 8mm 12mm 15mm,scale=0.542]{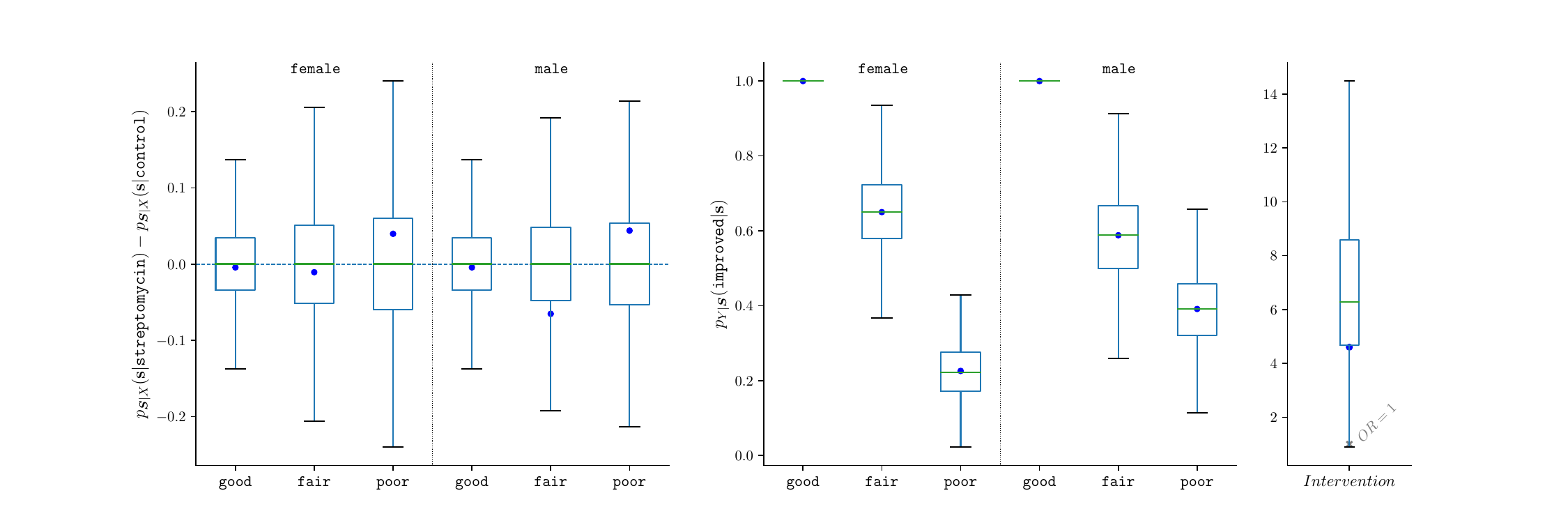}
    \caption{Box-plots on structural (dis)\textsc{p}arity, confounder \textsc{r}ealism and interventional $OR$ calculated over $10^6$ three-way contingency tables that were sampled from the \textsc{pr}-projection provided in Table~\ref{tb:Clinical1948}. Blue dots signify empirical estimates of the original study.}
    \label{fig:1948}
\end{figure}

Our goal is again to examine the effectiveness of \texttt{streptomycin} in a hypothetical setting where ratio \eqref{eq:1948:StructuralHeterogeneity} equals unity for all profiles in $\mathcal D_{\boldsymbol S}$. This automatically balances the conditional distribution $\prob f_{\boldsymbol S\vert X}$ in the two \texttt{treatment} groups via the observed marginal $\prob f_{\boldsymbol S}$, as opposed to methods that deal independently with each potential confounder~\cite{groenwold2011selection,pourhoseingholi2012control}.
Minimizing the $I$-divergence from $\prob f$ under \eqref{eq:PRconditions}, as before, we obtain the unique distribution $\prob q$ which satisfies structural \textsc{p}arity,
\equ{
\label{eq:1948:Parity}
 N q_{\boldsymbol S\vert X}(\boldsymbol s\vert x) = N  f_{\boldsymbol S}(\boldsymbol s) = \,
    \begin{array}{l|ccc}
    \renewcommand{\arraystretch}{1.1}
        & \texttt{good} & \texttt{fair} & \texttt{poor} \\\cline{2-4}
        \texttt{Female} & 8 & 20 & 31 \\
        \texttt{Male}   &  8& 17 &  23
    \end{array}
    \quads{\text{for}} x\in\{\texttt{streptomycin},\texttt{control}\}
}
with group prevalences $N\prob q_X(\texttt{control})=52$ and $N\prob q_X(\texttt{streptomycin})=55$, as in the original data; alongside the Realism of 
$N\prob q_{Y, \boldsymbol S} = N \prob f_{Y,\boldsymbol S}$ obtained by adding row-wise the latter two columns in Table~\ref{tb:Clinical1948}. 
Necessarily, the expected contingency table has fractional entries, e.g.\ $N q_{X,\boldsymbol S}(\texttt{control}, \texttt{Female}, \texttt{good}) = 8\cdot52/107\not\in\mathbb Z$ due to \textsc{p}arity \eqref{eq:1948:Parity}. 
Conceptually, this poses no problem, as $\prob q$ expresses the mere expectation of the confounding-free scenario and not a particular study over $N$ subjects.
The numerically determined $\prob q$ self-consistently leads to the hypothetical
\equ{
\label{eq:1948:corrected_Intervention}
Intervention_{\prob q} = \frac{Odds_{\prob q}(\texttt{improved} \vert \texttt{streptomycin})}{Odds_{\prob q}(\texttt{improved} \vert \texttt{control})\phantom{.........}} \approx 6.12
}
corresponding to an absolute risk reduction of more than $42\%$ in the \texttt{streptomycin} group.
Therefore, enforcing structural \textsc{p}arity seems to verify the conclusion of the original clinical trial  on the significant improvement due to \texttt{streptomycin} medication. 

Concerning three-factor effects, the 
$OR$ over the two groups for each patient profile $\boldsymbol s\in\mathcal D_{\boldsymbol S}$ from \eqref{eq:1948:s-Profiles} remains unaltered by the \textsc{pr}-projection:
\equ{
\frac{Odds_{\prob p}(\texttt{improved}\vert \texttt{control}, \texttt{gender},\texttt{baseline})\phantom{.........}}{Odds_{\prob p}(\texttt{improved}\vert \texttt{streptomycin}, \texttt{gender},\texttt{baseline})}
 = \,
    \begin{array}{l|ccc}
    \renewcommand{\arraystretch}{1.1}
        & \texttt{good} & \texttt{fair} & \texttt{poor} \\\cline{2-4}
        \texttt{Female} & \text{undefined} & 1/4 & 0 \\
        \texttt{Male}   &  \text{undefined}& 1/9 &  0
    \end{array}
    \quads{\text{for}}\prob p=\prob f,\prob q
    \nonumber
    ~.
}
On the other hand, the absolute risk reduction 
\equ{
\label{eq:1948:ARR}
ARR_{\prob p}=\prob p_{Y\vert X,\boldsymbol S}(\texttt{non-improved}\vert \texttt{control}, \boldsymbol s)
-
\prob p_{Y\vert X,\boldsymbol S}(\texttt{non-improved}\vert \texttt{streptomycin}, \boldsymbol s)
}
experiences a slight change
\equ{
ARR_{\prob f}
 \approx \,
    \begin{array}{l|ccc}
    \renewcommand{\arraystretch}{1.1}
        & \texttt{good} & \texttt{fair} & \texttt{poor} \\\cline{2-4}
        \texttt{Female} & 0 & 30\% & 41\% \\
        \texttt{Male}   &  0 & 46\% & 69\% 	
    \end{array}
    \quads{\text{vs.}}
    ARR_{\prob q}
    \begin{array}{l|ccc}
    \renewcommand{\arraystretch}{1.1}
        & \texttt{good} & \texttt{fair} & \texttt{poor} \\\cline{2-4}
        \texttt{Female} & 0 & 30\% & 44\% \\
        \texttt{Male}   & 0 & 48\% & 76\%
    \end{array}
    \nonumber
}
supporting the increased effectiveness of antibiotics in the hypothetical confounding-free scenario.

As a demonstration of the statistical nature of the \textsc{pr} approach, we 
generate from $\prob q$ one million contingency tables with $N=107$ patients each.
In Figure~\ref{fig:1948}, 
we indeed recover \textsc{p}arity and \textsc{r}ealism in expectation. 
Due to the estimation of fractional  cell counts (noted below \eqref{eq:1948:Parity}), no single contingency table with the marginals $N\prob f_{\boldsymbol S}$ and $N\prob f_X$  of the original study could exactly reproduce the expectation about \textsc{p}arity at the given sample size.
Because of the small sample size, the range of fluctuations is also large. Conversely, increasing the sample size $N$ would concentrate all cell counts from the generated contingency tables around the expected values that can be directly read off from $N\prob q$.
In the last box-plot, we estimate  the range of fluctuations at the given $N=107$ of the expected  effect size  \eqref{eq:1948:corrected_Intervention} of intervention in the bias-free scenario.

\subsection{Kidney stones}
\label{ssc:Kidney}

The clinical study~\cite{8e352ec3-c4c5-34a4-8ed5-ab6cf09d323f} was designed to compare different methods of
treating renal calculi.
The treatment was defined as \texttt{successful}, if stones were eliminated or reduced to less than $2mm$ after three months.
As it becomes clear from Table~\ref{tb:Kidney},
the $N = 985$ patients with renal calculi were distributed  over \texttt{treatment} groups\footnote{Additional treatment method \texttt{ureterolithotomy} and combined method \texttt{percutaneous nephrolithotomy} and \texttt{ESWL} are omitted from our investigation, because the original data contains unobserved confounder values.} in an unbalanced way:
\equ{
N \prob f_X = 
\begin{array}{r|r}
   \texttt{ESWL} & 328\\
   \texttt{nephrolithotomy/pyelolithotomy} & 231\\
    \texttt{percutaneous\,\,nephrolithotomy}  & 350\\
   \texttt{pyelolithotomy} & 76\\
\end{array}
~.
}
The same unbalanced trend applies for the proportion of patients with $s=\texttt{large}$ to $s=\texttt{small}$ stone size,
\equ{
\label{eq:Kidney:UnbalancedStones}
\frac{ f_{S\vert X}(\texttt{large} \vert x)}{ f_{S\vert X}(\texttt{small} \vert x)} \approx 
\begin{array}{r|r}
   \texttt{ESWL} & 0.61 \\
   \texttt{nephrolithotomy/pyelolithotomy} & 16.77 	 	\\
    \texttt{percutaneous\,\,nephrolithotomy}  & 0.30 \\
   \texttt{pyelolithotomy} & 1.45 	\\
\end{array}
}
within each \texttt{treatment} group $x$.
Taking the most populous group as reference, we comparatively estimate the effect size of $Intervention_{\prob f}$ as
\equ{
\label{eq:Kidney:Intervention}
\frac{Odds_{\prob f}(\texttt{successful}\vert x)\phantom{percutaneous..nephrolithotomy}}{Odds_{\prob f}(\texttt{successful}\vert \texttt{percutaneous\,\,nephrolithotomy})} \approx
\begin{array}{r|r}
   \texttt{ESWL} & 2.35 \\
   \texttt{nephrolithotomy/pyelolithotomy} & 0.54	 	\\
   \texttt{pyelolithotomy} & 1.13 	\\
\end{array}
~.
}
Based on such $OR$, 
$x=\texttt{pyelolithotomy}$ would not show a significant improvement ($p\text{-value}\approx 0.87$) compared to $x_0=\texttt{percutaneous\,\,nephrolithotomy}$.
In contrast, 
$x=\texttt{nephrolithotomy/pyelolithotomy}$ would appear 
significantly ($p\text{-value}\approx0.003$) inferior to reference treatment, while $x=\texttt{ESWL}$ 
significantly ($p\text{-value}\approx0.0004$) superior.
These conclusions are significantly ($p\text{-value}<10^{-11}$) obfuscated by  $S=\texttt{stone\,\,size}$,
\equ{
\frac{Odds_{\prob f}(\texttt{successful}\vert \texttt{small})}{Odds_{\prob f}(\texttt{successful}\vert \texttt{large})} \approx 3.51 	
}
generically increasing the chances of \texttt{successful} outcome in patients with $s=\texttt{small}$. This significant influence of \texttt{stone\,\,size} on the outcome had not been discovered and hence not accounted for in the original study design, as it becomes evident by \eqref{eq:Kidney:UnbalancedStones}.

\begin{table}[!t]
\renewcommand{\arraystretch}{1.05}
    \centering
    \begin{tabular}{ll|l||rr}
    $X=\texttt{treatment}$ & $S=\texttt{stone\,\,size}$ & $Y=\texttt{successful}$ & empirical $N\prob f$ & \textsc{pr}-projection\\\hline
     &  & \texttt{no} & 23 & 0.029433\\
    \texttt{extracorporeal\,shockwave} & \multirow{-2}{*}{\texttt{large}} & \texttt{yes} & 101 & 0.128444\\
     \rowcolor{\tablecolor}
    \texttt{lithotripsy} (\texttt{ESWL}) & & \texttt{no} & 4 & 0.003634\\
    \rowcolor{\tablecolor}
     & \multirow{-2}{*}{\texttt{small}} & \texttt{yes} & 200& 0.171484\\ \hdashline
    & & \texttt{no} & 64 & 0.032787\\
    \texttt{nephrolithotomy/} & \multirow{-2}{*}{\texttt{large}} & \texttt{yes} & 154 & 0.078401\\
    \rowcolor{\tablecolor}
    \texttt{pyelolithotomy} & & \texttt{no} & 1 & 0.010006\\
    \rowcolor{\tablecolor}
     & \multirow{-2}{*}{\texttt{small}} & \texttt{yes} & 12& 0.113324\\ \hdashline
    & & \texttt{no} & 25 & 0.052872\\
    \texttt{percutaneous} & \multirow{-2}{*}{\texttt{large}} & \texttt{yes} & 55 & 0.115594\\
    \rowcolor{\tablecolor}
    \texttt{nephrolithotomy} & & \texttt{no} & 36 & 0.026192\\
    \rowcolor{\tablecolor}
     & \multirow{-2}{*}{\texttt{small}} & \texttt{yes} & 234 & 0.160672\\ \hdashline
     &  & \texttt{no} & 7 & 0.005721\\
     & \multirow{-2}{*}{\texttt{large}} & \texttt{yes} & 38 & 0.030861\\
     \rowcolor{\tablecolor}
    & & \texttt{no} & 5 & 0.006868\\
    \rowcolor{\tablecolor}
    \multirow{-4}{*}{\texttt{pyelolithotomy}} & \multirow{-2}{*}{\texttt{small}} & \texttt{yes} & 26 & 0.033708\\ \hdashline
    \end{tabular}
    \caption{Three-way contingency table from the 1986 study on the effectiveness of different methods $x$ in treating renal calculi stratified by potential confounder $S$.}
    \label{tb:Kidney}
\end{table}

After eliminating structural differences among \texttt{treatment} groups,
\equ{
\frac{ q_{S\vert X}(\texttt{large} \vert x)}{ q_{S\vert X}(\texttt{small} \vert x)} = \frac{ f_{S}(\texttt{large})}{ f_{S}(\texttt{small})}=\frac{467}{518}\approx0.90
\quad\forall\,\,x\in\mathcal D_X
}
the \textsc{pr}-projection of $\prob f$ (last column in Table~\ref{tb:Kidney}) re-evaluates\footnote{Again, the Mantel-Haenszel method estimates instead approx.\ 3.37, 1.14 and 1.54 for the pooled $OR$ against \texttt{percutaneous\,\,nephrolithotomy} of \texttt{ESWL}, \texttt{nephrolithotomy/pyelolithotomy} and \texttt{pyelolithotomy}, respectively.} \eqref{eq:Kidney:Intervention} to
\equ{
\frac{Odds_{\prob q}(\texttt{successful}\vert x)\phantom{percutaneous..nephrolithotomy}}{Odds_{\prob q}(\texttt{successful}\vert \texttt{percutaneous\,\,nephrolithotomy})} \approx
\begin{array}{r|r}
   \texttt{ESWL} & 2.60 \\
   \texttt{nephrolithotomy/pyelolithotomy} & 1.28	 	\\
   \texttt{pyelolithotomy} & 1.47 	\\
\end{array}
~.
}
Given the hypothetical $OR$, we verify the significantly improving tendencies of $x=\texttt{ESWL}$ compared to $x_0$ from the original study.  At the same time, $x=\texttt{pyelolithotomy}$ which appeared non-significant in the original data due to small ($n=76$) size, exhibits a potential to be found significant in future studies, as the confounding-free scenario $\prob q$ raises its effectiveness. Finally, the significant inferiority of $x=\texttt{nephrolithotomy/pyelolithotomy}$ against $x_0$ previously detected by $\prob f$ is entirely reversed in the confounding-free scenario (cf.\ Simpson's paradox in Section~\ref{ssc:Simpson}). 

All in one analysis, we manage with one fictitious-study distribution $\prob q$ to exemplify all potential outcomes of a confounding-free scenario:  
\begin{itemize}
    \item verification of a previously  significant outcome
    \item enhancement of a previously non-significant outcome 
    \item complete overturning of the conclusion on the initial, confounded data
\end{itemize}
Based on the findings of the $\textsc{pr}$-projection,  research questions regarding those four therapeutic strategies could have been modified or even entirely reformulated in subsequent studies.

\subsection{COVID-19}
\label{ssc:COVID19}

Last but not least, we consider a more recent example from the \textsc{covid}-19 pandemic which demonstrates the power of our method to simultaneously  deal with multiple intervention groups $\vert\mathcal D_X\vert>2$ and mutliple confounder profiles $\vert\mathcal D_S\vert>2$. 
The data, already investigated by~\cite{9404149} in a similar context,  reports on $N=385\,925$ new \textsc{covid}-19 cases and deaths over the first wave until May 2020 in various countries stratified by $S=\texttt{age}$.
Out of $82\,798$ confirmed fatalities, the research question with $\mathcal D_Y=\{\texttt{survived},\texttt{deceased}\}$ concerns the mortality rate  
of the pandemic in each $X=\texttt{country}$ from
\equ{
\mathcal D_X = \left\{\texttt{Argentina},\texttt{Colombia}, \texttt{Italy},\texttt{Netherlands},\texttt{South\,\,Africa},\texttt{South\,\,Korea},\texttt{Switzerland} \right\}
~.
}
In a broader sense, the intervention in this application
pertains to the impact on the {mortality} rate of  various measures and living conditions aggregated by \texttt{country}. Essentially, we are interested to uncover what leads to discrepancies in the pandemic profile  among different \texttt{countries}.

\begin{figure}[t]
    \centering
    \includegraphics[scale=0.5,clip,trim=0.7cm 0.5cm 1cm 1.cm]{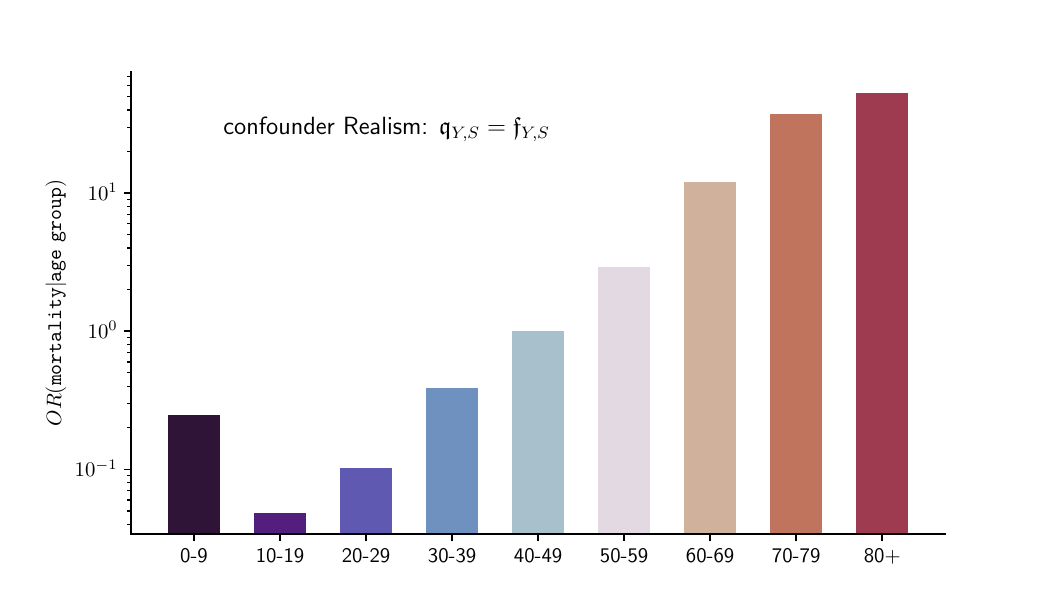}~
    \includegraphics[scale=0.5,clip,trim=1cm 0.7cm 1cm 1.5cm]{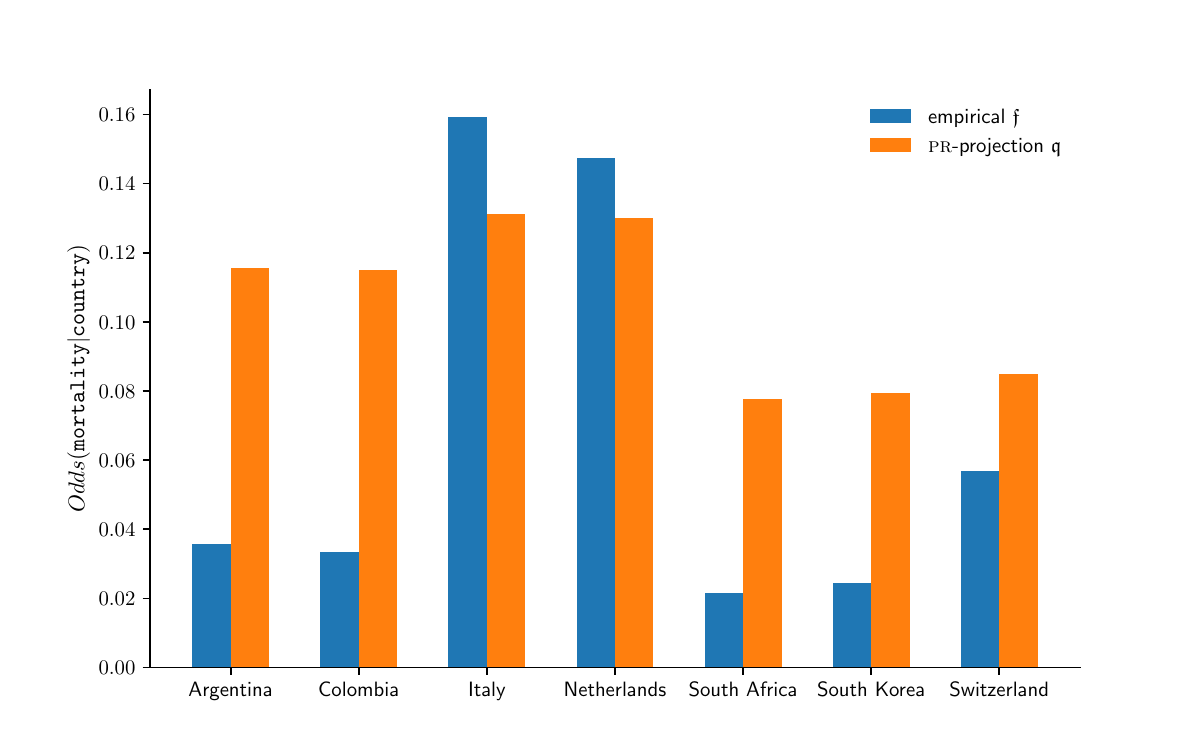}
    \caption{Left: global $OR$ for {mortality} due to \textsc{covid}-19 over the \texttt{age} groups as observed in the countries of the study ($s_0=\texttt{40-49}$ years as reference group).
    Right: {mortality} $Odds$ of \textsc{covid}-19 by \texttt{country} in 2020 empirically estimated by the official data $\prob f$ (blue) and re-estimated  by the \textsc{pr}-projection of $\prob f$ (orange). }
    \label{fig:Mortality-Age_Mortality-Country}
\end{figure}
As widely known and verified~\cite{RomeroStarkee006434}, \texttt{age} poses a risk factor for \textsc{covid}-19 mortality. In the present dataset, this is manifested by the noteworthy association of {mortality} rate  with the provided \texttt{age} groups in  Figure~\ref{fig:Mortality-Age_Mortality-Country} (left).
\begin{figure}[t]
    \centering
    \includegraphics[scale=0.71,clip,trim=1.cm 0.38cm 0.6cm 0.1cm]{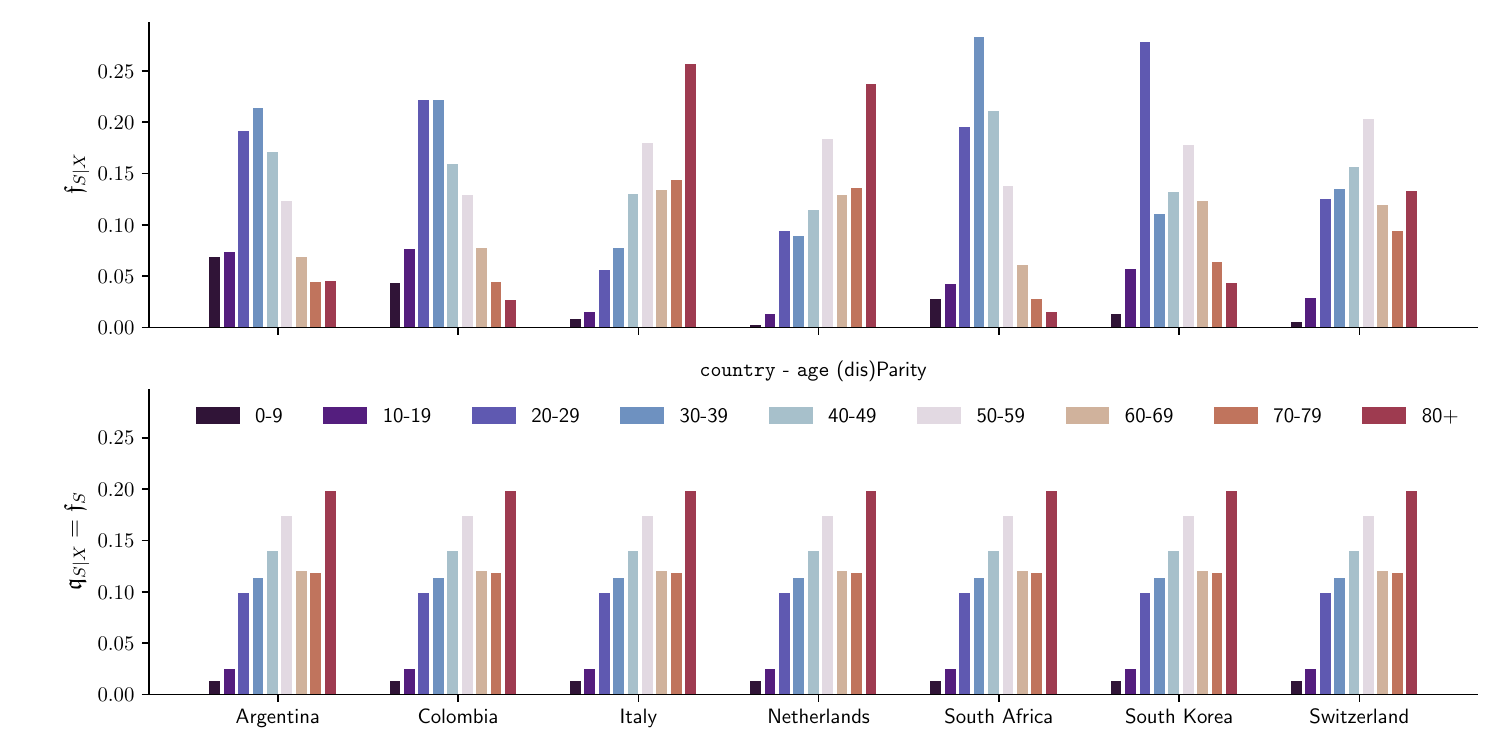}
    \caption{Upper: observed distribution of \textsc{covid}-19 cases over the \texttt{age} groups conditioned on the countries considered in the study. Lower: all considered countries share the same global distribution of cases over \texttt{age} groups in the hypothetical scenario of structural Parity.}
    \label{fig:Country-Age}
\end{figure}
However, not all countries exhibit the same distribution of \textsc{covid}-19 cases over \texttt{age} groups. This is exemplified by conditional $ f_{S\vert X}(\texttt{age}\vert\texttt{country})$ for all \texttt{age} groups from \texttt{0-9} to \texttt{80+} in the upper bar diagram of Figure~\ref{fig:Country-Age}. Most strikingly, \texttt{Italy} and the \texttt{Netherlands} seem to have a significantly higher case rate in \texttt{80+} years with \texttt{South\,\,Africa} exhibiting the lowest rate  in this age group. 
Empirically, the resulting {mortality} by \texttt{country} can be quantified from $\prob f$ via the
\equ{
\label{eq:COVID19:OddsMortality}
Odds_{\prob f}(\texttt{mortality}\vert\texttt{country}) = \frac{f_{Y\vert X}(\texttt{deceased}\vert\texttt{country})}{f_{Y\vert X}(\texttt{survived}\vert\texttt{country})}
}
which is depicted in the blue bar plot of Figure~\ref{fig:Mortality-Age_Mortality-Country} (right).

Establishing \textsc{p}arity in the \texttt{age} distribution of cases $q_{S\vert X}(\texttt{age}\vert \texttt{country})=f_S(\texttt{age})$ among each \texttt{country} leads to the hypothetical phenomenology of lower bar diagram~\ref{fig:Country-Age}. Since \texttt{Italy} had the highest incidence $ f_{X}(\texttt{Italy})\approx60\%$ in the dataset (followed by the \texttt{Netherlands} with almost $20\%$), the empirical marginal $ f_S(\texttt{age})$ and hence also the hypothetical  conditional $ q_{S\vert X}(\texttt{age}\vert \texttt{country})$ are dominated by the profile of $ f_{S\vert X}(\texttt{age}\vert \texttt{Italy})$.
This adjustment for structural \textsc{p}arity  alongside the confounder \textsc{r}ealism of \texttt{age} as risk factor depicted in Figure~\ref{fig:Mortality-Age_Mortality-Country} (left) lead to a new joint distribution $\prob q$, the \textsc{pr}-projection of $\prob f$, which re-estimates the  $Odds$ of \texttt{mortality} by \texttt{country} \eqref{eq:COVID19:OddsMortality} in the orange bar plot of Figure~\ref{fig:Mortality-Age_Mortality-Country} (right).

Adjusting for structural \textsc{p}arity, the $Odds$ for \texttt{mortality} significantly increase in countries where mostly younger \textsc{covid}-19 cases had been observed, while slightly decrease in European countries with a majority of older cases. 
After completely isolating the effect of \texttt{age} on \texttt{mortality} via the elimination of structural differences between \texttt{countries} regarding the profile of cases in Figure~\ref{fig:Country-Age}, we see a tendency of homogenizing the $Odds$ of \texttt{mortality} in each \texttt{country} (Figure~\ref{fig:Mortality-Age_Mortality-Country}, right). 
As anticipated, this alludes to \texttt{age} as a prominent risk factor for \textsc{covid}-19. 
However, the {mortality} rate does not become fully standardized in all investigated countries, as there exist further geographic-related factors  that could play a role in the course of a disease, such as local history of past pandemics, social attitudes against \textsc{covid}-19, differences in health-care systems, case-report and death protocols.

\section{Conclusions}

In this work, we have argued in favor of top-down approach from the three-way contingency table between response $Y$, independent variables $\boldsymbol X$ and confounders $\boldsymbol S$ towards the elimination of confounder-induced bias in the response to an intervention/exposure. 
After deriving the expectation about the de-biased contingency table, any metric like $OR$, $RR$ and $ARR$ that quantifies the influence of the intervention on the outcome
should be computed from the expected three-way table, otherwise it is usually meaningless. In that way, we avoid relying on external sources, special limits and model assumptions that generally perplex and --\,even most alarmingly\,-- could bias our analysis in an uncontrolled way beyond information theory.

In a model-agnostic exploration fully compatible with \totem, the minimization of $I$-divergence  under structural \textsc{p}arity proved to be a valuable tool. 
Through its unique solution, the $\textsc{pr}$-projection of $\prob f$  gives us the mathematical guarantee  that we do not unintentionally introduce any bias in our efforts to eliminate structural heterogeneity when going from empirical conditional $\prob f_{\boldsymbol S\vert\boldsymbol X}$ towards the hypothetical $\prob p_{\boldsymbol S\vert\boldsymbol X}=\prob f_{\boldsymbol S}$.
By doing so, we preserve --\,in expectation\,-- as much of original data's phenomenology as arithmetically feasible in a structurally homogeneous scenario. 
In latter scenario, any (non)-linear effect that might associate 
third-party variables 
 $\boldsymbol S$ to the  {independent} variables $\boldsymbol X$ has been, by construction, removed. 
In contrast to model-centric approaches that need to appeal to certain asymptotic limits presupposing conditions to be fulfilled by the data, the  information-theoretic methodology of the $\textsc{pr}$-projection always achieves structural \textsc{p}arity, as long as the suspected confounder profiles $\boldsymbol s$ are observed in all intervention groups $\boldsymbol x$.

The 
constrained $I$-divergence minimization in probability space leading to the \textsc{pr}-projection of the data 
elucidates that the optimization procedure of Section~\ref{ssc:iDiv}, which answers question~\ref{hypotheticalQ}, 
does not formally change with varying number of confounder manifestations and intervention groups, $\vert\mathcal D_{\boldsymbol S}\vert$ and $\vert\mathcal D_{\boldsymbol X}\vert$ respectively. 
Either we suspect multiple confounders (application~\ref{ssc:1948})  or deal with multiple  groups (examples~\ref{ssc:Kidney} and~\ref{ssc:COVID19}), the algorithmic procedure 
to obtain 
the
\textsc{pr}-projection, which unambiguously predicts metrics for the interventional effect size ($OR$, $RR$, $ARR$) free from confounder-induced bias, formally remains the same.  
Furthermore, the utility of an information-theoretic methodology in data analysis also encompasses privacy-preserving aspects.
Operating solely on contingency tables, namely summary statistics of microdata, which are provided in the original study, the \textsc{pr}-projection requires no additional input from individuals who  may be unwilling to share sensitive information beyond the study's scope.
To summarize the \textsc{pr} projection gives the mathematical guarantee to
\begin{itemize}[topsep=0.1pt,before=\vspace{1mm},after=\vspace{2mm}]
    \setlength\itemsep{0.1em}
    \item learn confounder statistics and the \textsc{r}ealism of confounders' influence on the outcome
    \item remove observed confounding by enforcing structural \textsc{p}arity among intervention groups
    \item avoid unintentionally introducing additional bias in doing so
    \item uniquely re-estimate the effect size of the intervention in the confounding-free scenario
    \item be applicable in multi-arm studies, also in presence of multiple confounders
\end{itemize}

In~\cite{loukas2023total}, it was argued that a metric variable can be phenomenologically regarded as categorical. More precisely, we define any variable over a finite, categorical domain $\mathcal D_\text{obs}$ of its observed-only values incorporating the physical ordering of metric values through observed moments, which constrain  the variable's distribution over $\mathcal D_\text{obs}$. 
In future work, we plan to naturally extend the \textsc{pr} approach of eliminating structural differences over count and metric variables as well as adjust the \textsc{pr}-projection for survival analysis. In addition, we shall revisit modeling of relationships between attributes, both as a pre-processing step to counteract vanishing cells in contingency tables and as a post-processing step,  after running the de-biasing to decide about significant trends in the structurally-homogeneous scenario. 


\subsection*{Acknowledgments}

We are thankful to Till Adhikary and Petros Marios Kitsaras for useful discussions as well as to Maria Chalkia for proofreading the manuscript. 

\renewcommand*{\bibfont}{\small}
\printbibliography

\end{document}